\documentclass[sigconf]{acmart}

\usepackage{amsmath,amsfonts,bm}

\def\eqref#1{equation~\ref{#1}}

\def\1{\bm{1}}

\DeclareMathAlphabet{\mathsfit}{\encodingdefault}{\sfdefault}{m}{sl}
\SetMathAlphabet{\mathsfit}{bold}{\encodingdefault}{\sfdefault}{bx}{n}

\newcommand{\sigmoid}{\sigma}

\usepackage[ruled,vlined]{algorithm2e}
\usepackage{color}
\usepackage{float}
\usepackage{import}
\usepackage{array}
\usepackage{bbm}
\usepackage{paralist}
\usepackage{subcaption}
\usepackage{amsmath} 
\usepackage{caption}
\usepackage{multirow}
\usepackage{makecell}
\usepackage{threeparttable}
\usepackage{graphicx} 
\usepackage{enumitem}
\setlist[itemize]{leftmargin=*}

\AtBeginDocument{%
  \providecommand\BibTeX{{%
    \normalfont B\kern-0.5em{\scshape i\kern-0.25em b}\kern-0.8em\TeX}}}

\setcopyright{acmcopyright}
\copyrightyear{2018}
\acmYear{2018}
\acmDOI{XXXXXXX.XXXXXXX}

\acmConference[Conference acronym 'XX]{Make sure to enter the correct
  conference title from your rights confirmation emai}{June 03--05,
  2018}{Woodstock, NY}
\acmPrice{15.00}
\acmISBN{978-1-4503-XXXX-X/18/06}

\graphicspath{./img}

\begin{document}


\title{Text2Bundle: Towards Personalized Query-based Bundle Generation}


\author{Shixuan Zhu}
\authornote{Both authors contributed equally to this research.}
\affiliation{%
  \institution{Tongji University}
  \city{Shanghai}
  \country{China}}
\email{2130768@tongji.edu.cn}

\author{Chuan Cui}
\authornotemark[1]
\affiliation{%
  \institution{Tongji University}
  \city{Shanghai}
  \country{China}}
\email{cuichuan@tongji.edu.cn}

\author{JunTong Hu}
\authornotemark[1]
\affiliation{%
  \institution{Tongji University}
  \city{Shanghai}
  \country{China}}
\email{1953697@tongji.edu.cn}

\author{Qi Shen}
\authornotemark[1]
\affiliation{%
  \institution{Tongji University}
  \city{Shanghai}
  \country{China}}
\email{1653282@tongji.edu.cn}

\author{Yu Ji}
\affiliation{%
  \institution{Tongji University}
  \city{Shanghai}
  \country{China}}
\email{2230779@tongji.edu.cn}

\author{Zhihua Wei}
\authornote{Corresponding author.}
\affiliation{%
  \institution{Tongji University}
  \city{Shanghai}
  \country{China}}
\email{zhihua_wei@tongji.edu.cn}

\renewcommand{\shortauthors}{XXX, et al.}



\begin{abstract}
Bundle generation aims to provide a bundle of items for the user, and has been widely studied and applied on online service platforms. 
Existing bundle generation methods mainly utilized user's preference from historical interactions in common recommendation paradigm, and ignored the potential textual query which is user's current explicit intention.
There can be a scenario in which a user proactively queries a bundle with some natural language description, the system should be able to generate a bundle that exactly matches the user's intention through the user's query and preferences.
In this work, we define this user-friendly scenario as Query-based Bundle Generation task and propose a novel framework Text2Bundle that leverages both the user’s short-term interests from the query and the user’s long-term preferences from the historical interactions. 
Our framework consists of three modules: (1) a query interest extractor that mines the user’s fine-grained interests from the query; (2) a unified state encoder that learns the current bundle context state and the user’s preferences based on historical interaction and current query; and (3) a bundle generator that generates personalized and complementary bundles using a reinforcement learning with specifically designed rewards. 
We conduct extensive experiments on three real-world datasets and demonstrate the effectiveness of our framework compared with several state-of-the-art methods.

\end{abstract}



\maketitle


%
%
\section{Introduction}

Bundles are ubiquitous in real-world scenarios, including the fashion outfit on the e-commerce platform Taobao, the music playlists on NetEase, the game packages on Steam, etc. 
A bundle is generally defined as a collection of items that are complementary or similar and can be consumed as a whole.   
Due to bundle's characteristics, the platforms are able to deliver highly relevant and satisfactory content to users while saving their time and effort, which brings more commercial benefits at the same time.
Typically, in bundle recommendation scenario, fixed or predefined bundles assembled by human expertise or non-personalized data mining methods are usually recommended \cite{DAM, BundleNet, HGCN}. 
Some advanced methods \cite{BRSteam, BGN, BuildBundle, BundleMCR} are able to recommend personalized bundles according to the interaction histories of users.
Besides, in search scenario, there is no method or paradigm for exploring the bundle-level response.

For the recommendation scenario without a user's proactive natural language input, the generated bundles based on the user's historical interaction may fail to meet some user specific demands.
Firstly, the user's intention is time sensitive, and the interaction based bundle generation methods cannot cope with the user's interest shift over time. 
Secondly, without the user defined instruction that specifies the context, attributes and constraints of the bundle, the bundle generation result would be uncontrollable.
Regarding the search scenario, there are also limitations in current paradigm to retrieve items. 
In this scenario, the user's textual query could be vague or abstract, but the search results could be excessively matched with the query such that the retrieved items can be homogeneous and only few of them would be selected by the user.
Meanwhile, the user may prefer to see a bundle-level result, but it is not supported by modern search systems.
Merely returning predefined bundles could not satisfy the user in terms of personalization and controllability.
Thereby, the user's query-based bundle generation is a significant task that needs to be tackled but not yet investigated.

\begin{figure}[t]
    \centering
    \includegraphics[width=\linewidth]{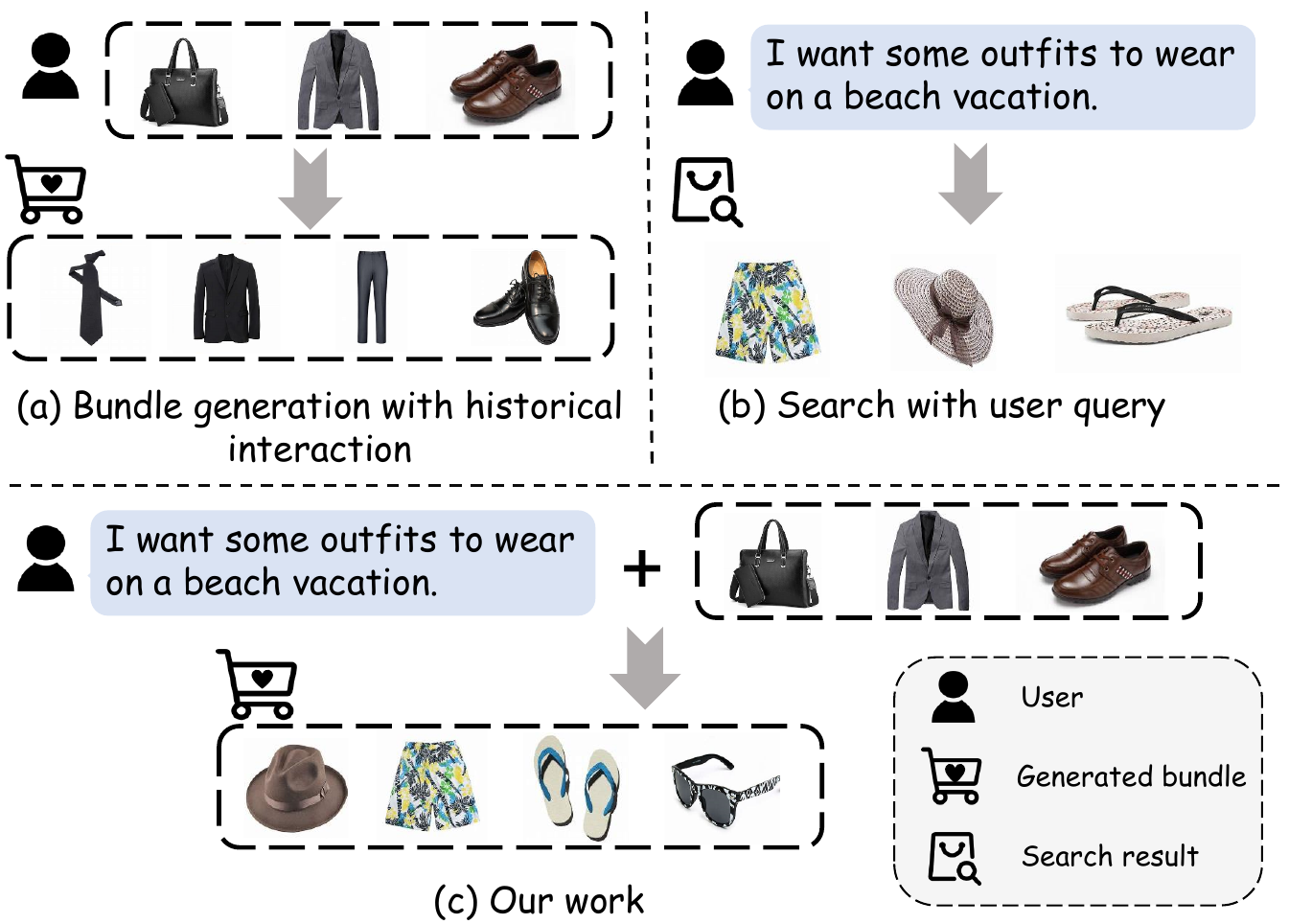}
    \captionsetup{font=small,labelfont=bf}
    \caption{An toy example of Query-based Bundle Generation scenario. }
    \label{fig:toy-example}
    \vspace{-0.3cm}
\end{figure}

In this paper, we mainly focus on this novel bundle generation task, which can be referred to as a Query-based Bundle Generation (QBG) process.
As shown in Fig. \ref{fig:toy-example}, when a user is looking for a clothing bundle with the descriptive text \emph{``I want some outfits to wear on a beach vacation.''}, the system would generate a bundle according to the user's historical interaction (Fig. \ref{fig:toy-example}.a) like \emph{handbag, suit and leather shoes}, the generated bundle would be things related like \emph{tie, suit, suit pant and leather shoe}, which are far away from the user's interest on beach vocation.
Or the system would search for matched items with the query (Fig. \ref{fig:toy-example}.b). But the user may need to select several items by himself, which is not user-friendly and there could be other desired items not listed.
Alternatively, in our work (Fig. \ref{fig:toy-example}.c), the recommender system can generate bundle for user based on the user’s intention inferred from the query and the user's preferences such as attribute \emph{men only} distilled from the interaction history simultaneously.

Previous studies on bundle generation rarely rely on the user's textual query but are simply based on user's interaction, focusing on techniques to generate bundles, for instance, using graph neural networks to represent bundles as graphs or using reinforcement learning to learn the optimal bundle combination strategy. 
In the new scenario that involves the user's query, the user's current short-term interests can be inferred from the query, along with the user's long-term interests reflected by the user's historical interaction. 
This suggests the new scenario could surpass the previous ones because both short-term and long-term user interests can be leveraged to generate personalized results for the user.
Spontaneously, to handle this scenario, we summarize the following three main research challenges:
\begin{itemize}[leftmargin=*]
    \item \textbf{How to mine the user's fine-grained interest?} 
    User's query is in text format, a straightforward way could be to encode the text by some pre-trained natural language models.
    However, the embedding of the whole sentence may fail to represent the rich fine-grained interests inside the query. 
    For example, directly encoding user's query in Fig. \ref{fig:toy-example} would possibly result in a vague interest like ``beach outfits'', but the user would also like the ``sunglasses'' to be included in the expected bundle but may not be captured by ``beach outfits''.

    \item \textbf{How to generate the personalized bundle?} 
    It is common that a user would have different preferences for various aspects such as colors, styles, and brands. These preferences can be inferred from the historical item interactions of the user. 
    To generate a personalized bundle, it is essential to utilize these preferences effectively. 
    Otherwise, the system may recommend an unsuitable bundle that matches the user’s current intention but deviates from the user’s preferences.

    \item \textbf{How to generate the qualified bundle? }
    Items in a qualified bundle should be complementary to each other. 
    For instance, a bundle with an \emph{iPhone} may also include a pair of \emph{earbuds}, but not an \emph{Android phone}. 
    Hence, we need to devise an effective method to measure the relations among items in a bundle.
    
\end{itemize}

To address these challenges, we propose a novel method named \textbf{Text2Bundle}, which employs Reinforcement learning and Large Language Model (LLM) for bundle generation.
Specifically, we utilize the Generative LLM to extract the fine-grained interests from the user's query, owing to its remarkable ability to infer user's intention with extend knowledge, which may be challenging or complicated for a conventional language model. 
To obtain the representation of the users and items efficiently, we derive both the ID and text embeddings from LightGCN pretraining procedure and natural language model encoding respectively.
Moreover, we introduce a unified state encoder to incorporate both short-term, long-term interests and the current bundle state from interaction and text modality.
Subsequently, our model selects the candidate item step by step based on the current state representation, with rewards considering the personalization, complementarity, and fine-grained interest coverage of current bundle.
Thus, the system can deliver a personalized and qualified bundle to meet user's real-time textual query. 
The main contributions of this work are as follows:
\begin{itemize}[leftmargin=*]
    \item We propose a novel and reasonable scenario named QBG, in which a user could proactively query a bundle with natural language descriptions, and the system would generate a bundle that exactly matches user's query and preferences.
    \item We propose a novel framework named Text2Bundle based on Reinforcement Learning that generates personalized and qualified bundles by leveraging the user's short-term and long-term interests.
    \item Extensive experiments on three bundle intention datasets are conducted to verify the effectiveness of our Text2Bundle framework.
\end{itemize}


\vspace{-0.2cm}
\section{Related Works}
\vspace{-0.05cm}
\subsection{Bundle Generation}
Bundle recommendation aims to recommend a bundle of items that are similar or complementary in content for a user to consume together. 
Existing works can be broadly classified into two categories: 1) ranking pre-defined bundles from the platform to users; 2) creating personalized bundles for users.

For bundle ranking, previous works often leverage additional information from user-item interactions and bundle-item affiliations.
For instance, DAM \cite{DAM} jointly models user-bundle and user-item interactions using multi-task neural networks to alleviate the scarcity of user-bundle interaction data.
BGCN \cite{BGCNBGGN} integrates the relations among users, bundles, and items into a heterogeneous graph and applies graph neural networks \cite{GNNBook2022} to learn the complex relationships.
Moreover, CrossCBR \cite{crosscbr} models the cooperative associations between the two perspectives using cross-view contrastive learning.

The key challenge of generating personalized bundles is how to produce a set of items that not only meet the user’s personalized demands but also have internal coherence, which requires accurate modeling of user interests as well as the compatibility of items within the bundle.
BGN \cite{BGN} models bundle generation as a structured prediction problem and uses determinantal point processes (DPPs \cite{DPPs}) to generate high-quality and diversified bundles.
However, this approach models the bundle as a sequence, which may fail to capture the relationships between distant items.
BGGN \cite{BGCNBGGN} represents the bundle as a graph and employs graph neural networks to generate the graph, which learns the structural information and high-order item-item relationships.
BYOB \cite{BuildBundle} formulates the problem as a combinatorial optimization problem over a set of candidate items and applies a policy-based deep reinforcement learning algorithm to solve it.

However, In real-world recommendation scenarios, such as e-commerce platforms, users may express their own intention through a query and expect to receive personalized bundles that satisfy both their intention and historical preferences.
Previous works do not explicitly model this textual user intention, which may lead to inadequate modeling and suboptimal results.

\vspace{-0.05cm}
\subsection{RL in Recommendation}
Reinforcement learning (RL) is a machine learning method that learns optimal policies by interacting with the environment and maximizing reward signals and can be applied to various domains such as games, robot control, and recommendation systems.
In this method, agents receive observations from the environment, make decisions based on the current policy, and update the policy according to the reward signals.

Compared to other recommendation methods, RL-based recommendation systems have the capability to handle the dynamics of sequential user-system interactions by adjusting actions according to successive feedback received from the environment. 
Additionally, RL takes into account the long-term user engagement with the system, allowing for a better understanding of users' preferences.
RLUR \cite{RLUR}  utilizes reinforcement learning to enhance user retention, which models the problem as an infinite horizon request-based Markov Decision Process, with the goal of minimizing the accumulated time interval of multiple sessions, thereby improving the app open frequency and user retention.
HRL-Rec \cite{integration2} aims to model user preferences on both item and channel levels, in order to jointly recommend heterogeneous items from multiple channels and satisfy users' personalized and diversified information needs.


Bundle generation can be formulated as a combinatorial optimization problem, but the number of possible item combinations increases exponentially with the number of items, and traditional algorithms cannot solve this problem in polynomial time.
Moreover, capturing the relationships between items within a bundle is crucial for effective bundle generation.
To address these challenges, BYOB \cite{BuildBundle} obtains optimal item combinations through a policy-based deep reinforcement learning algorithm and designs several item-level reward signals to address the data sparsity problem.

However, this approach has some limitations.
Firstly, it assumes that the size of the generated bundle is fixed, which cannot suit real scenarios. 
Secondly, it does not adequately model the relationships among users, bundles, and items.
The policy network only utilizes mean-pooling and fully connected layers to model the relationships between user and selected items for a bundle, which neglects the higher-order relationships among them.
Additionally, a previous study \cite{KnowActionSet} has shown that in the cases where actions are interdependent, optimal action decisions should consider other available actions. 
BYOB only considers compatibility modeling between actions and existing items within the bundle, neglecting the modeling of dependencies among actions themselves.

\vspace{-0.05cm}
\subsection{Large Language Model in Recommendation}
Recently, the LLMs such as ChatGPT \cite{chatgpt} and LLaMa \cite{touvron2023llama} emerges to show their strength in various tasks involving natural language such as question answering, text summarization, translation, and more, due to training on massive amounts of text data.
The capabilities like Chain-of-thought \cite{wei2022chain}, instruction following \cite{ouyang2022training} and in-context learning make LLMs much more powerful than traditional natural language models, 
which would improve the explainability and efficacy of the recommender systems.

P5 \cite{p5} proposes a unified and shared conditional language generation framework which integrates several recommendation tasks, and shows zero-shot generalization ability for novel personalized prompt and new items in unseen domains.
ChatRec \cite{chatrec} utilizes the LLM's in-context learning ability to establish connections between users and items, enabling interactivity and explainability enhanced multi-round recommendations.

While these methods mainly focus their scenarios to integrate LLMs to item recommendation or conversational recommendation, there is no LLMs usage on bundle recommendation for now.
Besides, we leverage LLMs for intention decomposition as an enhancement to bundle generation process, and it is not an end-to-end usage of LLMs different from aforementioned methods.


%

%
%
%

\begin{figure*}
    \centering
    \includegraphics[width=0.95\textwidth]{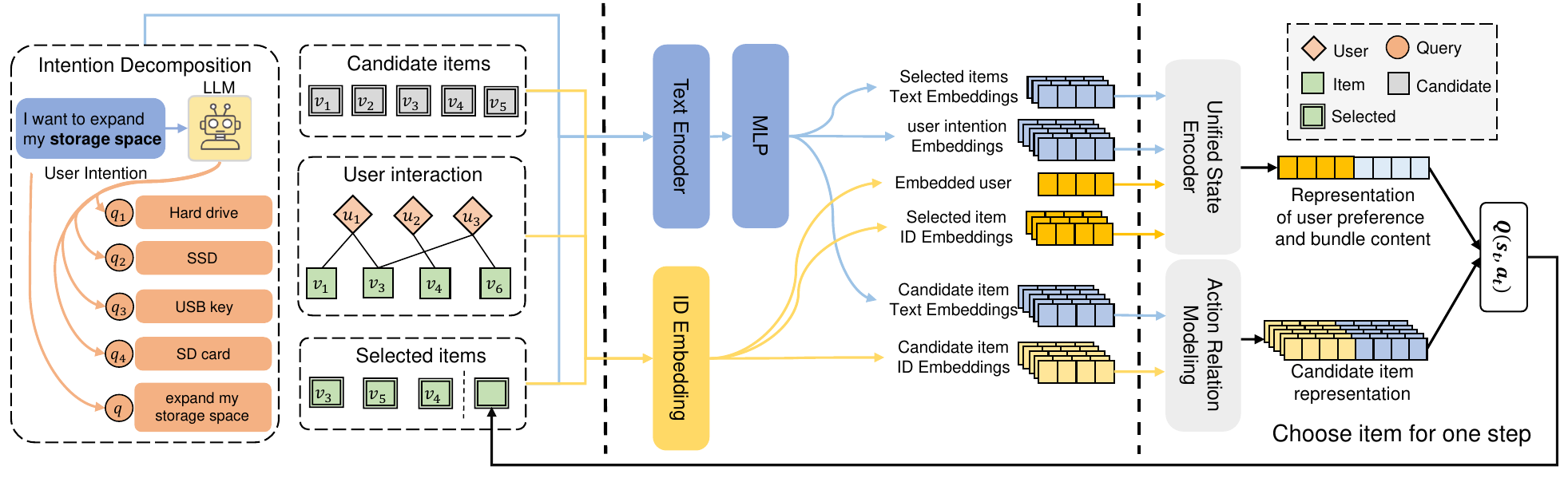}
     \caption{The overview of Text2Bundle. The key rationale of our framework is to add suitable next item into bundle based on current state composed of user preferences and bundle content, then update the state for next item prediction. It first decomposes the intention query into multiple intention instances via LLM, and constructs candidate item pool by a simple Action Selection Strategy. Then in each step, it employs the Unified State Encoder to distill state representation from multiple information, and utilizes Action Relation Modeling to acquire candidate action representation. Finally, the item is chosen by these two representations to match user's preference and textual query, and be complementary to current selected items.}
    \label{fig:framework}
\end{figure*}

\vspace{-0.1cm}
\section{Query-based Bundle Generation}\label{sec:define}
\vspace{-0.05cm}
\subsection{Definition}
\vspace{-0.05cm}
In this work, we define the user, item, and bundle as $u \in \mathcal{U}$,  $v \in \mathcal{V}$, and $b \in \mathcal{B}$. 
The QBG scenario can be formulated as: for a user $u$ with interaction history $\mathcal{V}_u$ and a raw textual query $q$, generate a bundle $b^{*}_{u,q} = \{v_1,v_2,\dots,v_K\}$ which meets the user's intention in $q$ and preferences in $\mathcal{V}_u$, where $K$ denotes the variational length of the bundle for each query session. 
Note that we omit the subscript in $b^{*}_{u,q}$ in the following discussion for simplicity. 

\vspace{-0.1cm}
\subsection{Framework}
Inspired by BYOB \cite{BuildBundle}, we formulate the query-based bundle generation as a combinatorial optimization problem over the items in the candidate pool. 
We then transform the problem into a Markov Decision Process (MDP), in which an item is added to the bundle iteratively, and the process can be mainly regarded as a selection decision from candidate items.
The MDP can be formulated as RL formulation $\mathcal{M} = <\mathcal{S}, \mathcal{A}, P, r, \gamma  >$:
\begin{itemize}[leftmargin=*]
\item \textbf{State space $\mathcal{S}$.} For each step $t$, we define the state $s_t \in \mathcal{S}$ in the context of user $u$ as $\{u, \mathcal{V}_u, p^t, b^t\}$.
Where $b^t$ is current bundle of selected items, and $p^t$ is the candidate pool with selected items excluded.
For the initial state $s_0$, $p^0$ includes all the candidate items and an additional end action $a_{end}$, and $b^0$ is an empty set. 

\item \textbf{Action space $\mathcal{A}$.} An action $a \in p^t$ denotes the matched item selected from the candidate pool $p^t$ based on current bundle $b^t$ and user's short-term and long-term interests, or the end action $a_{end}$ to stop the generation process.

\item \textbf{Transition $\mathcal{P}$.} The transition between states can be formulated as $s_{t+1} = (u, \mathcal{V}_u, p^{t+1} = p^{t} \setminus \{a\}, b^{t+1}=b^t \cup \{a\})$. 
Where a selected item $a$ is added to current bundle $b^t$ and removed from the candidate pool $p^{t}$. 
When the bundle size reaches the maximum bundle size $L$ defined as a hyperparameter, or the end action $a_{end}$ is selected, the bundle generation process will be stopped.

\item \textbf{Reward $r$.} The reward function is to guide the item selection to make the generated bundle qualified, it will be elaborated on in following sections. 

\item \textbf{Discount factor $\gamma$.} The discount factor balances the trade-off between immediate and future rewards. 
\end{itemize}
The framework is to learn a bundle generation policy $\pi(a|s;\theta)$, to select the proper action $a$ based on current state $s$ by maximizing the expected reward over evaluation episodes, where the $\theta$ is the parameters of the policy network.

\vspace{-0.3cm}
\section{Methodology}\label{sec:method}
\vspace{-0.1cm}
The main objective of this work is to generate a bundle that meets the user’s satisfaction and qualification criteria. 
To achieve this, the user’s short-term and long-term interests are crucial. 
Furthermore, the reinforcement learning policy and reward design are indispensable for a qualified bundle that contains harmonious items. 
In this section, we present the details of the user’s short-term and long-term interest and bundle state modeling, and the reinforcement learning policy.

%

\vspace{-0.2cm}
\subsection{Intention Decomposition}\label{sec:decomp}
\vspace{-0.05cm}
As stated earlier, we tend to infer the fine-grained intention instances from the textual query $q$, while an end-to-end training process can be very difficult. 
Thus, we harness the remarkable ability of Generative LLMs for intention inferring and knowledge extending \cite{kojima2022large} to facilitate the extraction of user’s fine-grained interests from the query $q$, which would be challenging for a conventional model.

\noindent \textbf{Prompt design \;}\label{sec:prompt-design}
For a specific task, Generative LLM is able to generate corresponding answers with a proper prompt provided.
Therefore, We design an instruction prompt \emph{
``I will list a description of a bundle which is an item set, please help to extract 4-5 possible entities implied inside. 
Please insert a "|" between each entity. Here is an example: \emph{Query: } I want to expand my storage space; \emph{Answer: } Hard drive|SSD|USB key|SD card'' } and with an input query as we discussed before in Fig. \ref{fig:toy-example}, the detailed intentions will be output with "|" separated for better data processing. 
Thus, we can decompose user's query $q$ to the textual intention instances $\mathcal{Q} = \{{q}_1,{q}_2,...,{q}\}$. 
Note that we append the original query string ${q}$ to the collection because the original semantics in the query can be kept in case of the potential sub-optimal intention decomposition.  




\vspace{-0.1cm}
\subsection{State Modeling of User Preferences and Bundle Content}\label{sec:user-interest-modeling}
To select a proper candidate item for current bundle, the user's long-term and short-term interests and the content of the bundle should be taken into account for modeling.
To model the relation of each candidate item with the user and items in current bundle comprehensively, the naturally textual information of user's input and items, and the rich user-item interaction data should be adequately utilized.
To this end, we uniformly apply Transformer \cite{transformer} to encode the user interests and current bundle from the text level and interaction level.

\subsubsection{ID Embedding Encoder}\label{sec:id embedding}
The interactions between user and items in bundle are inherently sparse which would lead to insufficient training by learning the ID embeddings from scratch.
We introduce a pre-trained LightGCN \cite{he2020lightgcn} in our work to address this issue, as detailed in Sec. \ref{sec:dataset}. Eventually, we can obtain the effective ID embeddings of users and items.
For a specific user $u$ and item $v_i$, they can be represented as $\mathbf{e}_{ID}^u $, $ \mathbf{e}_{ID}^{v_i} \in \mathbf{R}^{d}$, where $d$ is the dimension of the user and item embeddings.

\subsubsection{Text Embedding Encoder}
We employ the RoBERTa \cite{liu2019roberta}, a pre-trained text embedding model, to encode the title corpora of the items as embedding: $\mathbf{e}_{text}^{v_i} = \text{RoBERTa}(title_{v_i})$.
For the decomposed textual user intention instances or the query itself, we use the same encoding method to obtain their text embeddings: $\mathbf{e}_{text}^{q_j} = \text{RoBERTa}(q_j)$, where $q_j \in \mathcal{Q}$.

\subsubsection{Unified State Encoder}\label{sec:unified}
In order to obtain the unified state,  we expect the textual semantic embeddings can be fused with the ID embeddings, and a shared multi-layer perception (MLP) was employed to transform text embeddings to the space of ID embeddings. 
In detail, we apply the MLP on the text embeddings $\mathbf{e}_{text} \in {\mathbf{R}}^{d_t}$ of historical item $v_i \in \mathcal{V}_u$, intention instance $q_j$, or selected item $v_k \in b^t$ to obtain the transformed embeddings $\mathbf{e}_{text'} \in \mathbf{R}^{d}$, where $d_t$ and $d$ are the embedding sizes.

Then the short-term from user's current intention instances, long-term interest from historical interactions and the items in current bundle, in both ID and text modality, can be fused by a Transformer to a representation of current state of the user $u$ and bundle $b^t$.
We concatenate all these embeddings $\{\mathbf{e}_{text'}^{q_j}\}$,$\{\mathbf{e}_{text'}^{v_i}\}$, $\{\mathbf{e}_{text'}^{v_k}\}$, $\mathbf{e}_{ID}^{u}$, $\{\mathbf{e}_{ID}^{v_i}\}$, $\{\mathbf{e}_{ID}^{v_k}\}$ as the input set, through the Transformer we can obtain the refined representations respectively.
Note that besides original position embedding in Transformer, we also add the corresponding type embedding $\mathbf{e}_{type} \in \mathbf{R}^{d}$ into input embedding, to indicate above $6$ different input embedding types.
By this Transformer Encoder, we model the cross modal information and relations among items in the current bundle, and fuse the long-term and short-term interests, also mine possible missing items of the current bundle in the context of the user's query.
Moreover, we employ average pooling on all output text embedding to obtain the text state embedding $\mathbf{e}_{text'}^{state} \in \mathbf{R}^{d}$, and the same for the state intention embedding $\mathbf{e}_{ID}^{state}\in \mathbf{R}^{d}$.
By concatenating these two interest embeddings on their embedding dimension, we can obtain the final state embedding $\mathbf{e}^{state} \in \mathbf{R}^{2d} $ of current state $s_t$:
\begin{equation}
\setlength{\abovedisplayskip}{2pt}
\setlength{\belowdisplayskip}{2pt}
\begin{aligned}
    \mathbf{e}^{state} = \text{Concat}(\mathbf{e}_{text'}^{state}, \mathbf{e}_{ID}^{state})    
\end{aligned}\end{equation}

\vspace{-0.1cm}
\subsection{Bundle Generation Policy}\label{sec:agent}
Upon obtaining the user interest, we employ the deep Q-learning network (DQN) \cite{DQN} to perform the policy learning in QBG scenario.
Due to the overestimation bias in original DQN, we employ the double DQN \cite{DDQN} to copy a target network $Q'$ as a periodic from the online network to train the model following \cite{zhou2020interactive,Unicorn}.
The policy network receives the embeddings of current state and candidate action as input and produces the value $\mathbf{Q}(\mathbf{s}_t,\mathbf{a}_t)$ for each candidate action:
\begin{equation}\setlength{\abovedisplayskip}{2pt}
\setlength{\belowdisplayskip}{2pt}\begin{aligned}
    \mathbf{Q}(\mathbf{s}_t,\mathbf{a}_t) = \mathbf{e}^{state} (\mathbf{E}^{act}_{a_t})^T
\end{aligned}\end{equation}
where $\mathbf{E}^{act}_{a_t} \in \mathbf{R}^{2d}$ is the embedding of candidate action $a_t$ in action embedding matrix  $\mathbf{E}^{act}$ (See Eq. \ref{Eq:action_embedding}), generated by following two steps: \emph{Action Selection Strategy} and \emph{Action Relation Modeling}.

\subsubsection{Action Selection Strategy} \label{sec:act_selection}
As the performance of the policy learning could be degraded by a large action space \cite{Unicorn} that is essential to be dealt with, i.e., massive unfiltered candidate items, in QBG scenario. 
To this end, we propose a simple but effective recall method. 
Specifically, we first calculate the similarity \cite{sentencebert} between text embeddings of the user's query and all the items to filter out items below a specific similarity threshold $T_{sim}$ to get a recall candidate set $\mathcal{V}_{recall}$.
Furthermore, we incorporate an FM \cite{fm} recommendation model with the user and item features to get top $R$ items from $\mathcal{V}_{recall}$ as the candidate set $\mathcal{V}_{cand} = p^0$ for current user's query.

\subsubsection{Action Relation Modeling}
In QBG scenario, the candidate items in the action set could be interactive and relational with each other.
For instance, a one-piece swimsuit ranked in the top $1$ of candidates would be good for a beach outfit.
Meanwhile, beach shirts and pants ranked in the top $2$ and $3$ are complementary and would be also favored by the user.
If a one-piece swimsuit was selected, beach shirts and pants should not be in the bundle anymore, and vice versa.
Thus, we need to take the combination composability of candidate items into account.
To achieve this, we employ a Transformer to obtain the candidate actions embeddings by modeling the relations among them:
\begin{equation}\setlength{\abovedisplayskip}{2pt}
\setlength{\belowdisplayskip}{2pt}\begin{aligned}
    \mathbf{E}^{v_{cand}}=\text{Concat}(\mathbf{E}^{v_{cand}}_{ID}, \text{MLP}(\mathbf{E}^{v_{cand}}_{text}))
\end{aligned}\end{equation}
\begin{equation}\setlength{\abovedisplayskip}{2pt}
\setlength{\belowdisplayskip}{2pt}\begin{aligned}
    \mathbf{E}^{act}=\text{TRM}_{act}(\mathbf{E}^{v_{cand}}) \cup \{\mathbf{e}^{a_{end}} \}
\label{Eq:action_embedding}
\end{aligned}\end{equation}
where $\mathbf{E}_{ID}^{v_{cand}} \in \mathbf{R}^{(|p^t|-1) \times d}$ and $\mathbf{E}^{v_{cand}}_{text} \in \mathbf{R}^{(|p^t|-1) \times d_t}$ are the ID and text embeddings of the candidate items, $\mathbf{e}^{a_{end}} \in \mathbf{R}^{2d}$ is the learnable embedding of the end action $a_{end}$.

\subsubsection{Reward Settings}\label{sec:reward}
We expect the generated bundle $b^{*}$ to be similar to the target bundle $\hat{b}$ which can be regarded as the label in the dataset, and items in $b^{*}$ should be more complementary. 
Hence, we use the precision value (see details in Sec. \ref{Sec:metrics}) of $b^{*}$ and $\hat{b}$ as the main reward $r_{main}$.

\noindent \emph{Personalization and complementarity.}
However, during the agent's exploration, only a hit can result in a positive precision reward.
This leads to a very sparse reward and potentially slow learning progress or even failure to learn.
To mitigate this issue, we apply the reward shaping process. 
Specifically, we use the user-item match score, calculated by the dot-product operation of user and item embeddings from the pre-trained LightGCN (see more details in the Sec. \ref{sec:dataset}), as an auxiliary precision reward $r_{per}$ to indicate the satisfaction of user $u$ with the action item $a_t$.
To ensure the complementarity of items in the bundle, we calculate the dot product of the embeddings of selected items in $b^t$ and the action item $a_t$ as the complementarity reward $r_{comp} = \mathop{\mathrm{MEAN}} \limits_{v_k \in b^t} {(\mathbf{e}^{v_k}_{text}(\mathbf{e}^{a_t}_{text})^T)}$, where $\mathbf{e}^{v_k}_{text}, \mathbf{e}^{a_t}_{text} \in \mathbf{R}^{d_t}$.

\noindent\emph{Coverage of fine-grained interest.}
Considering all the fine-grained interests extracted from user's query might be preferably covered, we introduce an entropy-like reward to ensure this. 
In detail, for each item $v_k$ in the generated bundle $b^t$, we calculate $\mathcal{I}(v_k) = \mathop{\mathrm{argmax}}\limits_{q_j \in \mathcal{Q}} {  \mathbf{e}^{v_k}(\mathbf{e}^{q_j})^T }$ to determine which intention instance this item may belong to, measured by the similarity scores between item and each intention instance.
Thus, we can obtain the count of items $\mathcal{C}(q_j)=|\{ v_k| \mathcal{I}(v_k) = q_j\}|$ included for each intention instance $q_j$.
The reward can be represented as:
\begin{equation}
\setlength{\abovedisplayskip}{2pt}
\setlength{\belowdisplayskip}{2pt}
\begin{aligned}
    r_{cover}(b^t, \mathcal{Q}) = - \sum_{q_j \in \mathcal{Q}} p(\mathcal{C}(q_j)|b^t)\log{p(\mathcal{C}(q_j)|b^t)}, 
\end{aligned}\end{equation}
\begin{equation}\setlength{\abovedisplayskip}{2pt}
\setlength{\belowdisplayskip}{2pt}
\begin{aligned}
    p(\mathcal{C}(q_j)|b^t) = \frac{\mathcal{C}(q_j)}{|b^t|} 
\end{aligned}\end{equation}
When the distribution of $q_j$ is more likely even, the $r_{cover}(b^t, \mathcal{Q})$ will be larger to guide more coverage on all the fine-grained interests.

Finally, we aggregate all these rewards into a reward to guide the RL training process:
\begin{equation}\setlength{\abovedisplayskip}{2pt}
\setlength{\belowdisplayskip}{2pt}\begin{aligned}
    r = \omega \cdot r_{main} + r_{per} + r_{comp} + r_{cover},
\end{aligned}\end{equation}
where $\omega$ is a hyperparameter to determine the weight of the main reward.

\subsubsection{Recommendation Pre-Training}\label{rec-train}
Considering the sample efficiency and instability problems of RL, it is difficult and time-consuming to train the bundle generation policy from scratch.
Therefore, we first pretrain the unified state encoder part (Sec . \ref{sec:unified}) with a recommendation task based on semi-synthetic data.
Here, we aim to pave the way for the bundle generation task under RL framework by the supervised learning of user-item recommendation and bundle completion recommendation.
In detail, we expect the distance between current generation state representation $\mathbf{e}^{state}$ and the target items, i.e. the items in the target bundle $\hat{b}$, to be smaller than others.
To achieve these goals, we employ pairwise Bayesian Personalized Ranking (BPR) loss \cite{bpr} as follows:
\begin{equation}
\setlength{\abovedisplayskip}{2pt}
\setlength{\belowdisplayskip}{2pt}
\begin{aligned}
    \mathcal{L}_{p}=\sum_{(u,q,b^t, v_{+},v_{-})\in \mathcal{D}_{pre}} -\ln \sigmoid(cos(\mathbf{e}^{state},\mathbf{e}^{v_+})-cos(\mathbf{e}^{state},\mathbf{e}^{v_-})),\,
\end{aligned}\label{eqn:loss_bpr}
\end{equation}
where $\mathcal{D}_{pre}$ denote the constructed training set of item pairs,  and $\sigmoid$ is sigmoid function.
$\mathcal{D}_{pre}:=\{(u,q,b^t,v_{+},v_{-})|v_{+}\in \hat{b}\setminus b^t, v_{-}\in \mathcal{V}\setminus (\hat{b}\cup b^{t})\}$, where $b^{t}$ is the synthetic bundle for $t$-th step, $\hat{b}$ is the target bundle of query $q$, $v_{+}$ is the ground truth item to be added but uncovered in $b^t$.
Here, $b^t$ is composed of $t$ items randomly selected from the target bundle $\hat{b}$, where $t$ is ranged from $0$ to $|\hat{b}|-1$, simulating the entire completion steps in subsequent RL learning. 

Based on the above pre-training process, we further train the policy module and unified state encoder jointly.

\vspace{-0.15cm}
\section{Experiments}
\vspace{-0.05cm}


%


In this section, we conduct experiments on the QBG scenario to evaluate the performance of our method compared with other state-of-the-art (SOTA) models \footnote{Our code and data will be released for research purposes.}.


\newcommand{\tabincell}[2]{\begin{tabular}{@{}#1@{}}#2\end{tabular}}  
\begin{table}[htbp]
\vspace{-0.15cm}
\setlength{\abovecaptionskip}{0cm}  
\setlength{\belowcaptionskip}{-0.15cm} 
\captionsetup{font=small,labelfont=bf}
    \caption{Statistics of datasets used in experiments.}
    \label{tab:dataset}
    \centering
    \small    
    \begin{tabular}{lrrr}
    \toprule
    Statistic& Clothing & Electronic & Food  \\
    \midrule
    No. of items  & 4,487 & 3,499 &3,767  \\
    No. of users  & 965 & 888 &879 \\
    No. of bundles &1,910 & 1,750 &1,784 \\
    No. of Intents &1,466 & 1,422 &1,156 \\
    \midrule
    No. of user-bundle &1,912 &1,753 &1,785 \\
    No. of user-item &6,326 &6,165 &6,395 \\
    \midrule
    Avg. of bundle size & 3.31 & 3.52 & 3.58  \\
    \bottomrule
    \end{tabular}
    \vspace{-0.3cm}
\end{table}


\vspace{-0.15cm}
\subsection{Dataset} \label{sec:dataset}
\vspace{-0.05cm}

We conduct experiments on \emph{Bundle Intent dataset} \cite{BundleIntentDS} containing user intentions (i.e., the textual query) to evaluate our proposed method.
Bundle intention refers to user's intuitive feeling of a bundle.
In order to obtain bundle intention, the authors of Bundle Intent Dataset design a crowd-sourcing task to label potential bundles and corresponding intentions hidden in the user session from the three domains (Electronics, Clothing, and Food) extracted from the Amazon dataset \cite{AmazonDS}, thereby construct a high-quality bundle dataset with bundle intentions. 
We utilize the entire Amazon datasets (the data source of Bundle Intention Dataset) to pre-train the LightGCN \cite{he2020lightgcn} model, and extract the user-item embeddings (Sec. \ref{sec:id embedding}) contained in the Bundle Intent Dataset for further training.

\vspace{-0.15cm}
\subsection{Baseline Models}
\begin{itemize}[leftmargin=*]
\item{\textbf{BPR}} \cite{bpr} is a traditional recommendation method based on Bayesian analysis to rank the items according to the user's implicit feedback.
\item{\textbf{BGN}} \cite{BGN} regards bundle generation as a structured prediction problem and utilizes detrimental point processes to generate a high-quality and diversified bundle list.
\item{\textbf{Bunt}} \cite{BundleMCR} is a multi-round conversation recommendation method for bundle generation. 
\item{\textbf{BYOB}} \cite{BuildBundle} is a competitive personalized bundle generation model within RL paradigm, which generates the bundle with the guidance of multiple rewards. 
\item{\textbf{LLM4B}} is our designed LLM-based bundle generation baseline, following ChatRec \cite{chatrec}. It first recalls top-$30$ items ordered by their semantic similarity \cite{sentencebert} with the query. Then we choose some items from them as the bundle with the prompt like: \emph{``I want you to recommend a bundle of items based on user query and user's historical interactions, user query:\{user query\}.The historical records include the item title and description. You are encouraged to learn user preference from the interacted items:\{historical interactions\}. Here is a list of items that user is likely to pick \{candidate item list\}. Please select some complementary items that meet both user query and preference to form a bundle, separated by commas between the items''}, where \emph{\{user query\}, \{historical interactions\} and \{candidate item list\}} indicate the input query, titles of historical items and candidate items respectively.
\end{itemize}

We note that most of the original methods above are incompatible with our QBG scenario.
To achieve a fair comparison, we make several modifications to these baselines. 
For $\textbf{BGN}^\dagger$, we consider the top first bundle in its generated bundle list as the result. 
For $\textbf{Bunt}^\dagger$, we fuse the text embedding of the user's query to its transformer input and use the bundle generated in the first round for comparison.
We add the text embedding to the representation of the state and apply an auxiliary reward which is the textual similarity between user's query and title of the action item, to the model $\textbf{BYOB}^{\dagger}$.

\begin{table*}[t]
    \centering
    \setlength{\abovecaptionskip}{0cm}  
\setlength{\belowcaptionskip}{-0.1cm} 
\captionsetup{font=small,labelfont=bf}
    \caption{Experimental results (\%) of different models in three metrics on three datasets. The bold number indicates the improvements over the best baseline (underlined) are statistically significant ($p \textless 0.01$) with paired t-tests.}
    \label{tab:overall}
    \begin{tabular}{p{2.6cm}<{\centering}p{0.55cm}<{\centering}p{0.7cm}<{\centering}p{0.7cm}<{\centering}p{0.01cm}p{0.55cm}<{\centering}p{0.7cm}<{\centering}p{0.7cm}<{\centering}p{0.01cm}p{0.55cm}<{\centering}p{0.7cm}<{\centering}p{0.7cm}<{\centering}}
    \toprule
    \multirow{2}{*}{ \bfseries Models}& \multicolumn{3}{c}{ \bfseries Clothing }& & \multicolumn{3}{c}{\bfseries Electronic }& &\multicolumn{3}{c}{\bfseries Food} \\
    \cline{2-4}
    \cline{6-8}
    \cline{10-12}
    &Pre &Rec &F1 &&Pre &Rec &F1 &&Pre &Rec &F1\\
   
    \midrule

    $\text{BPR}$   &35.6 &18.3 &24.1 
    &  & 32.2 & 18.1 &23.2    &&37.2  &19.2 &25.3   \\

    $\text{BGN}^\dagger$   &47.3 &17.1 &25.1 
    &  &46.9 &18.4 &26.4   &&48.0  &18.9 &27.1  \\

    $\text{Bunt}^\dagger$   &48.9 &18.8 &27.1 
    && 53.1  &17.1 &25.9  &&49.5  &\underline{19.8} &\underline{28.2}  \\
    
    $\text{BYOB}^\dagger$   &\underline{65.1} &\underline{19.5} &\underline{30.0} 
    &  & \underline{61.8} & \underline{19.7} &\underline{29.9}    && \underline{62.9}  & 17.9 &27.9  \\
    
    $\text{LLM4B}$  &43.6 &18.4 &25.8 
    & & 41.9 &16.9 &24.0 && 44.1 &18.1 &25.7 \\
    
    \midrule
    Text2Bundle (Ours) &{\bfseries 81.8}& {\bfseries 20.5} &{\bfseries 32.8}   &&{\bfseries 81.3} &{\bfseries 21.1} &{\bfseries 33.5} &&{\bfseries 80.7} &{\bfseries 20.3} &{\bfseries 32.4}\\
    \bottomrule
    \end{tabular}
    \vspace{-0.4cm}
\end{table*}

\vspace{-0.2cm}
\subsection{Evaluation Metrics}\label{Sec:metrics}
For bundle generation, some position-sensitive ranking related metrics, such as Normalized Discounted Cumulative Gain (NDCG) and Mean Reciprocal Rank (MRR), may not provide an accurate evaluation, since bundle is an unordered collection, and the order of items within a bundle should not influence the evaluation metrics.
To evaluate the general performance of bundle generation, similar to \cite{BuildBundle}, we use the Precision, Recall, and F1-score to measure the similarity between generated bundle and target bundle.

Notably, our model employs token $a_{end}$ to generate bundles with adaptive lengths, while previous works could only generate bundles with predefined fixed bundle size $K$. 
Therefore, evaluation metrics based on fixed bundle size, such as Pre@K and Rec@K, may not be directly applicable to this work. 
For baselines, we generated bundles with a fixed size of $K=5$, aligned with our max bundle size setting, based on the common bundle length observed in the dataset. 
By employing this standardized setting, the performance of our proposed model and other baselines could be compared fairly.

\textbf{Precision}:
The precision measures the quality of generated bundle, which could be formulated as follows:
\begin{equation}
\setlength{\abovedisplayskip}{2pt}
\setlength{\belowdisplayskip}{2pt}
\begin{aligned}
    Precision=\frac{1}{|\mathcal{D}|}\sum_{(u, q, b^*, \hat{b}) \in \mathcal{D}}\frac{|b^*\cap \hat{b}|}{|b^*|} 
\end{aligned}\label{eqn:pre}
\end{equation}
where $\mathcal{D}$ represents the evaluation dataset, $|\cdot|$ represents the size of set, $b^*$ and $\hat{b}$ represents the generated bundle and ground-truth bundle of user $u$ and query $q$. 

\textbf{Recall}:
Recall measures the ability of a model to predict all target items of the ground-truth bundle.
Mathematically, Recall can be defined as follows:
\begin{equation}
\setlength{\abovedisplayskip}{2pt}
\setlength{\belowdisplayskip}{2pt}
\begin{aligned}
    Recall=\frac{1}{|\mathcal{D}|}\sum_{(u, q, b^*, \hat{b}) \in \mathcal{D}}\frac{|b^*\cap \hat{b}|}{|\hat{b}|} 
\end{aligned}\label{eqn:recall}
\end{equation}

\textbf{F1-score}:
F1-score is a metric that combines both Precision and Recall, providing a balanced assessment of the bundle generation performance, defined as $F1 = 2\cdot{\frac{\text{Precision}\cdot \text{Recall}}{\text{Precision}+\text{Recall}}}$.

All these metrics range from $0$ to $1$, with higher values indicating better performance.

\vspace{-0.15cm}
\subsection{Implementation Details}\label{sec:details}
We implement the proposed method based on Pytorch.
We randomly divide the samples into training, validation, and test parts with the ratio of $8:1:1$ for each dataset.
The ID embedding dimension $d$ is set as $320$, and the text embedding dimension $d_t$ is set as $384$.
We employ the Adam optimizer to train the bundle recommendation module.
The weight of the main reward $\omega$ is set to be $3.0$.
Max bundle size $L$ is set to be $5$. 
During the action selection process (Sec. \ref{sec:act_selection}), we set the similarity threshold $T_{sim}$ to filter items to $0.3$, and the
candidate items size $R = |p^0|$ to $50$.
For the Transformer encoder in Unified State Encoder and Action Relation Modeling, we employ a single encoder layer with two attention heads.
To transform textual embeddings to the space of ID embeddings, we employ a two-layer MLP with dimensions $[64, 64]$.
We set the learning rate to $5e-5$ and conduct experiments for a maximum of 1000 epochs on the three datasets to obtain the final results.

We employ ChatGLM \cite{chatglm}, an open-source dialogue language model based on General Language Model (GLM) framework, to make intention decomposition.
For the baseline LLM4B model, we employ GPT-$3.5$ \cite{chatgpt} to ensure the performance of bundle generation.

We employ the tianshou \cite{tianshou} framework to implement the reinforcement learning algorithm. 
The experience replay memory size is $20,000$, the sample batch size is $256$, and the discount factor $\gamma$ is set to be $0.99$.
We re-run all experiments three times with different random seeds and report the average performance.

\vspace{-0.2cm}
\subsection{Overall Comparison}\label{sec:exper}
To demonstrate the comprehensive performance of the proposed model, we have adapted state-of-the-art (SOTA) methods for bundle generation to suit the Query-based Bundle Generation (QBG) scenario.
By comparing our model with these established methods, we aim to provide a thorough evaluation of its effectiveness and superiority in generating high-quality bundles.
As shown in \autoref{tab:overall}, we can obtain the following observations.
\textbf{(1)} In the QBG scenario, our proposed model Text2Bundle outperforms significant performance improvements over all baseline models across three datasets, particularly in terms of the precision metric.
The enhanced performance of our model can be attributed to the effective fusion of long-term user preferences and short-term query information. 
This integration enables our model to consider both the historical behavior and immediate textual query, resulting in a more accurate understanding and representation of their interests, allowing for a comprehensive capture of user interests.
In particular, compared to the BYOB, our model incorporates unified state modeling and conducts fine-grained interest decomposition for user queries, which contributes to a deeper understanding and characterization of user profiles.
\textbf{(2)}  In most cases, text-based models ($\text{Bunt}^\dagger$, $\text{BYOB}^\dagger$, LLM4B and Text2Bundle) perform significantly better than models that solely rely on interaction information (BPR and $\text{BGN}^\dagger$), indicating the effectiveness of incorporating textual information in our scenario.
The improved performance of text-based models suggests that leveraging textual data, in addition to interaction information, contributes to a better understanding and modeling of user preferences and interests, leading to more accurate and personalized recommendations in our scenario.
\textbf{(3)} For $\text{Bunt}^\dagger$, $\text{BYOB}^\dagger$ and LLM4B, as can be seen, the performance order can be summarized as $\text{BYOB}^\dagger >\text{Bunt}^\dagger >\text{LLM4B}$.
The performance disparity between $\text{BYOB}^\dagger$ and $\text{Bunt}^\dagger$ can be primarily attributed to Bunt's disregard for the complementary among items, resulting in sub-optimal bundle generation.
Despite the fact that the $\text{LLM4B}$ method ignores user-interaction data, its performance closely approximates that of $\text{Bunt}^\dagger$, which suggests the strong potential of LLM-based models for bundle generation.



\begin{table}[t]
    \setlength{\abovecaptionskip}{0cm}  
\setlength{\belowcaptionskip}{-0.1cm} 
\captionsetup{font=small,labelfont=bf}
    \caption{Results (\%) of the Ablation Study. }
    \label{tab:ablation_study}
    \footnotesize
    \centering
    \begin{tabular}{p{2.95cm}<{\centering}p{0.4cm}<{\centering}p{0.45cm}<{\centering}p{0.45cm}<{\centering}p{0.01cm}p{0.4cm}<{\centering}p{0.45cm}<{\centering}p{0.45cm}<{\centering}}
    \toprule
    \multirow{2}{*}{\bfseries Models }& \multicolumn{3}{c}{\textbf{Clothing}}&& \multicolumn{3}{c}{\textbf{Electronic}}\\
    \cline{2-4} \cline{6-8}
    &Pre &Rec &F1  &&Pre &Rec &F1\\
    \midrule
    $-$Generative LLM &76.7& 18.1 &31.9 &&  76.0& 19.4& 30.9\\
    \midrule
    $-$Unified State Encoder &45.6& 17.5 &27.3 &&  39.0& 19.0& 27.3\\
    \midrule
    $-$Action Relation Modeling&66.9& 18.4 &31.3 &&  68.6& 19.7& 30.6\\
    \midrule
    Only Text &40.1& 18.3 & 27.0  &&  49.0& 18.7& 29.1 \\
    Only Id &52.9& 18.8 & 29.9  && 59.9& 19.7&31.9 \\
    \midrule
    Text2Bundle (Ours) &81.8& 20.5 &32.8 &&  81.3& 21.1& 33.5\\
    \bottomrule
    \end{tabular}
    \vspace{-0.4cm}
\end{table}

\vspace{-0.15cm}
\subsection{Ablation Study}\label{sec:ablation}
In this section, we conduct some ablation studies on the proposed model to investigate the effectiveness of some designs. 
Specifically, we develop the following variants:
\textbf{(1)}``-Generative LLM'': A variant of Text2Bundle that eliminates the Generative LLM for intention decomposition.
Instead, during the process of the Unified State Encoder, the original textual query is utilized as a substitute.
\textbf{(2)}``-Unified State Encoder'': Replacing the Unified State Encoder with a Multi-Layer Perceptron (MLP).
\textbf{(3)}``-Action Relation Modeling'': A variant of Text2Bundle which removes the Action Relation Modeling module.
\textbf{(4)}``Only Text'' and ``Only Id'': The Unified State Encoder module and the Action Relation Modeling module only take ID Embeddings (or Text Embeddings) as input, while another modality inputs are masked. 

As shown in Table~\ref{tab:ablation_study}, the model performance decreases significantly while removing the intention decomposition ability provided by Generative LLM and just using user's original query for intention inferring.
This result highlights the crucial role played by intention decomposition in the model's performance. 
The ability to decompose the textual query into multiple instances that can capture user's fine-grained intentions proves to be pivotal in effectively understanding and representing user interests.
Meanwhile, The absence of Unified State Encoder and Action Relation Modeling also leads to a sharp decline in generation performance. 
Among them, the precision value after removing Unified State Encoder is almost halved, which is even worse than the performance of only use textual embedding or only use id embedding.  
It shows that the alignment of representations in different semantic spaces is helpful and necessary for state modeling.


\begin{table}[t]
    \setlength{\abovecaptionskip}{0cm}  
\setlength{\belowcaptionskip}{-0.1cm} 
\captionsetup{font=small,labelfont=bf}
    \caption{Results of the Human evaluation. }
    \label{tab:human}
    \footnotesize
    \centering
    \begin{tabular}{p{1.3cm}<{\centering}p{0.4cm}<{\centering}p{0.85cm}<{\centering}p{0.85cm}<{\centering}p{0.01cm}p{0.4cm}<{\centering}p{0.85cm}<{\centering}p{0.85cm}<{\centering}}
    \toprule
    \multirow{2}{*}{\bfseries Models }& \multicolumn{3}{c}{\textbf{Clothing}}&& \multicolumn{3}{c}{\textbf{Electronic}}\\
    \cline{2-4} \cline{6-8}
    &Comp &Align-Q &Align-U  &&Comp &Align-Q &Align-U\\
    \midrule
    $\text{BYOB}^\dagger$ &3.50 &3.42 & 3.23 && 3.74  & 3.88 &3.53 \\
    LLM4B & 3.86 &3.93  & 3.12 &&3.89  & 3.85 & 3.17 \\
    Text2Bundle &4.25  &4.37  &3.51  &&4.18  & 4.24 &3.64 \\
    \bottomrule
    \end{tabular}
    \vspace{-0.3cm}
\end{table}

\vspace{-0.2cm}
\subsection{Human Evaluation}

Besides automatic evaluation, we also conduct human evaluations to compare the performance of the baselines and our methods on Clothing and Electronic dataset.
The experiment involves $12$ post-graduate volunteers who evaluate $40$ samples each, with each sample being evaluated by $3$ different volunteers.
In each sample, volunteers are presented with a user historical items and current query, along with the generated bundle by $\text{BYOB}^\dagger$, LLM4B, and Text2Bundle. 
Volunteers are instructed to answer following questions: \textbf{1}). ``Is bundle complementary?''  \textbf{2}).``Does bundle align with the given query?'' \textbf{3}).``Does bundle match the user's profile?'', and evaluate three results in each sample on a scale of $1-5$, with $1$ being ``worst'' and $5$ being ``best'',  from these three evaluation views respectively.
The results are shown in Table \ref{tab:human}, where the metric \emph{Comp}, \emph{Align-Q}, and \emph{Align-U} indicate the average evaluation scores of above three questions correspondingly.
On all three evaluation aspects, our Text2Bundle outperforms other two competitive baselines, which is generally consistent with the result of automatic evaluation.
However, different from the result of the automatic evaluation, annotators are inclined to give LLM4B results higher scores than $\text{BYOB}^\dagger$ when evaluating bundle complementarity and query-level alignment. 
This could be explained that automatic evaluation computes the similarity between the generated bundle and the target bundle strictly, since some items that may be qualified and substituted but are not in the target bundle are still considered poor choices.  
While these items may be deemed reasonable by the annotator, and account for a greater proportion of LLM4B's result than $\text{BYOB}^\dagger$'s.
Meanwhile, this may also be due to the inconsistency between annotator and reality user, resulting in the bias of the expected bundle.

\begin{figure}[t]
    \centering
    \includegraphics[width=0.40\textwidth]{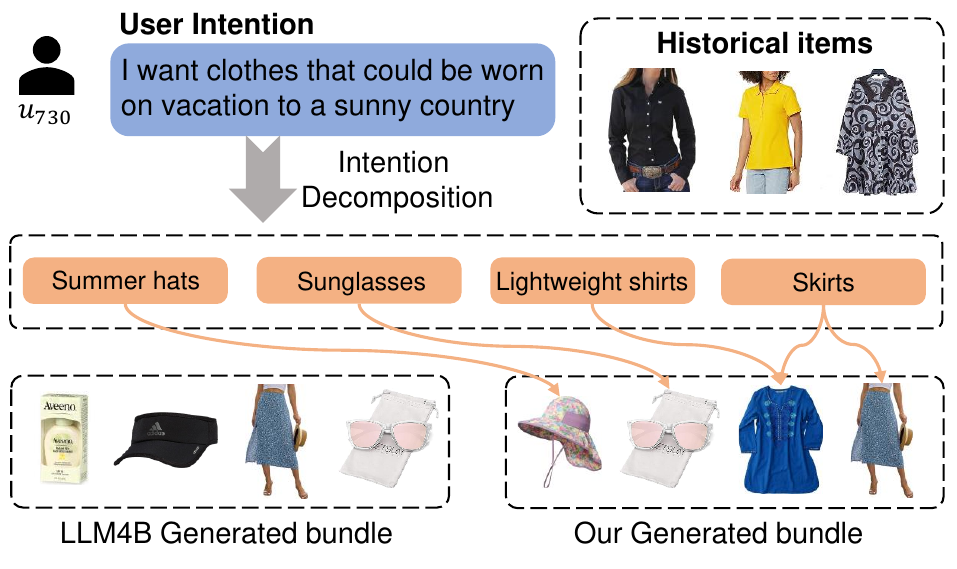}
    \caption{Snapshots of generated bundles using Text2Bundle and LLM4B.}
    \label{fig:case_study}
    \vspace{-0.6cm}
\end{figure}

\vspace{-0.05cm}
\subsection{Case Study}\label{sec:case}
\vspace{-0.05cm}
To intuitively present the personalized generative results of our model, we present a recommendation case in Fig. \ref{fig:case_study}. 
The user (with \emph{ID 730}) has a history of interacting with predominantly female tops and skirts. 
Her textual input is expressed as \emph{``I want clothes that could be worn on vacation to a sunny country''}.
During the initial phase, the system decomposes the user's textual query into four finer-grained intention instances: \emph{summer hats, sunglasses, Lightweight shirts, and skirts}.

Subsequently, the system performs several steps of bundle generation, continuing until it encounters end action $a_{end}$.
The final four items form a complete bundle, and there exists a correspondence (orange curve in Figure) between the generated bundle and the instance derived from the intention decomposition.
Additionally, the generated bundle aligns well with the user's preference.

As a comparison, LLM4B choose the \emph{sunscreen} for the bundle, which does not match user's requirement of \emph{wearable} specified in her query. 
This suggests that LLM4B may tend to select relevant but not precisely fitting items, while our model avoids this phenomenon by intention decomposition.
The intention decomposition in our model extracts user's fine-grained interest as intention instances, which serve as anchors to emphasize the main content of bundle generation.
Meanwhile, the bundle generated by LLM4B does not contain \emph{tops} items, indicating insufficient complementarity. 
In contrast, our intention decomposition module may alleviate this issue through the explicit intention instance for better coarse to fine planning and instance-level measurement for complementarity.
Furthermore, during the RL process, the bundle state modeling, the action relation modeling, and multiple rewards $r_{comp}$, $r_{cover}$ also enhance the complementarity among the items within the bundle.

\vspace{-0.15cm}
\section{Conclusion}
\vspace{-0.1cm}
In this paper, we first present a novel recommendation scenario, query-based bundle generation, where system can generate a personalized bundle for user's query. 
For this scenario, we propose a new method named Text2Bundle.
Specifically, it generates the bundle item by item, based on the fine-grained user intention instances and long-term user preference.
Extensive experiments on three datasets verify the effectiveness and superiority of Text2Bundle in the proposed scenario.
The new scenarios and the proposed methods open a new door to more realistic and general bundle generation and potentially to other areas such as conversational recommendation.

\bibliographystyle{ACM-Reference-Format}
\bibliography{ref}


\appendix
\newpage

\newpage
\setcounter{section}{0}

\begin{table}[!htp]
    \centering
    \small
    \setlength{\abovecaptionskip}{0cm}  
    \setlength{\belowcaptionskip}{-0.1cm} 
    \captionsetup{font=small,labelfont=bf}
    \caption{Some snapshots of generated bundles using Text2Bundle and LLM4B, where \textbf{H}, \textbf{Q},\textbf{I}, \textbf{L} and \textbf{T} indicate the \textbf{H}istory items, user \textbf{Q}uery, intention \textbf{I}nstances, \textbf{L}LM4B and \textbf{T}ext2bundle respectively.}
    \label{tab:allcase}
    \begin{tabular}{p{0.1cm}<{\centering}p{8cm}}    

    \toprule
    \bfseries H & Kids Sandal, Mickey Rolling BackpackHide, Baby Girl's 2-Pack Pants\\ 
    \bfseries Q  & Some items I would take on holiday with me  \\
    \bfseries I & Tops, Bottoms, Outerwear, Footwear, Accessories \\
    \bfseries L & Pocket Hoodie, Fashion Sneaker, Heels Mid Calf Boots, Polarized Sunglasses\\
    \bfseries T & Baby-Girls Swimsuit, Women's Fashion Sneaker, Women Dress Long Sleeve Top Blouse\\
    
    \midrule
    \specialrule{0em}{1.5pt}{1.5pt}
    \midrule

    \bfseries H & Men's Classic Fleece Pant, DILLIGAF Pin, Women's Chloe Mule\\ 
    \bfseries Q  & A set of items to buy for a baby shower.\\
    \bfseries I & Baby clothing, Baby accessories, Baby footwear, Baby blankets, Baby hats\\
    \bfseries L & Baby Pants, Baby-Girls Swimsuit, Sweet Kids Jacket, Kids Sandal, Women's Nursing Bra\\
    \bfseries T & Women's Wedge Sandal, baby shower hat, Baby Newborn Romper\\ 
    
    \midrule
    \specialrule{0em}{1.5pt}{1.5pt}
    \midrule

    \bfseries H & Women's Bra, Slingback Sandal, Belted Maxi Dress, Jasmine Slingback Sandal\\ 
    \bfseries Q  & Accessories for a pretty lady!\\
    \bfseries I & Jewelry, Handbags, Scarves, Hair accessories, Sunglasses\\
    \bfseries L & 18k Gold-Plated Necklace,  Women's Slingback Sandal, Polarized Sunglasses, Round Dress Watch\\
    \bfseries T & 18k Gold-Plated Necklace, Polarized Sunglasses, Seatbelt Belt, Metal earrings \\ 
    
    \midrule
    \specialrule{0em}{1.5pt}{1.5pt}
    \midrule

    \bfseries H &Amazon Kindle Paperwhite, DVD+/-RW Drive, UVB Cable Ties, LCD Display, Samsung NX1000 Camera\\ 
    \bfseries Q  & I want to upgrade personal computer desktop\\
    \bfseries I & CPU, GPU, Memory (RAM), SSD, Power Supply Unit\\
    \bfseries L &SATA Controller Card, 10.1-Inch Tablet, LCD Monitor, Graphics Card, Power Supply\\
    \bfseries T &Samsung Galaxy Desktop Dock, keyboard, SATA Controller Card, USB flash drive\\ 
    
    \midrule
    \specialrule{0em}{1.5pt}{1.5pt}
    \midrule

    \bfseries H &Wireless Shutter Release Remote Control, Sandisk memory card, Keyboard Cover\\ 
    \bfseries Q  & Something for streaming gameplay on Twitch or other streaming sites\\
    \bfseries I & High-performance Gaming PC , Capture Card, Webcam, Microphone, Streaming Software\\
    \bfseries L & Keyboard and Laser Mouse,  LCD Monitor, Dynamic Microphone, Game Capture\\
    \bfseries T &LCD Monitor, Video Adapter, GPU, Microphone\\ 
    
    \midrule
    \specialrule{0em}{1.5pt}{1.5pt}
    \midrule

    \bfseries H &Quick Charger, Power Amplifier, Color E-reader, Media PlayerPrimer\\ 
    \bfseries Q  & Camera and its equipments\\
    \bfseries I &Camera Body, Lenses, Tripod, Camera Bag, Lens Filters\\
    \bfseries L &Digital Camera, Nikon Battery, Sandisk 8GB Memory Card, Wireless Shutter Release Remote Control\\
    \bfseries T &Wireless Shutter Release Remote Control, Digital to Analog Convertor, Nikon Battery, Digital Camera\\ 
    
    \midrule
    \specialrule{0em}{1.5pt}{1.5pt}
    \midrule

    \bfseries H & 10.1-Inch Tablet, Battery-Free BlueEye Mouse\\ 
    \bfseries Q  & Something for Building a computer and entertainment to go with it. \\
    \bfseries I &Computer Components, Monitor or Display, Keyboard and Mouse, Speakers or Headphones\\
    \bfseries L & 10.1-Inch Tablet, LCD Monitor, Keyboard and Laser Mouse, TV Wall Mount Bracket, Capsule Speaker\\
    \bfseries T &USB controller card, Capsule Speaker, LCD Monitor, Keyboard, Av to Tv-rca Cable\\ 
    

    

    


    \bottomrule
    \end{tabular}
\end{table}

\section{Additional Results}
As shown in \autoref{tab:allcase}, we present additional $7$ snapshots of generated bundles based on Text2Bundle and LLM4B to demonstrate the specific performance of our model.

\section{Further Discussion}
In this section, we briefly compare our proposed \emph{Query based Bundle Generation} scenario to other bundle recommendation related scenarios to highlight the novelty of our work.

Traditional bundle generation methods generate bundles mainly based on user history interactions. 
Here, the user history interactions and the item in the bundle are still in the same modality. 

One scenario similar to our QBG is \emph{Conversational Bundle Generation}, as Bunt designed for, which initials the bundle based on user's query and updates the bundle based on user's multi-round feedback. 
While Bunt is built based on a simple user simulator, where the user's query and feedback are simple attributes instead of natural language.
That means, the ability of Bunt‘s instruction following is limited to the attribute-level, and cannot achieve a more fine-grained and scalable natural language understanding.
Besides, it focuses more on multi-turn conversations, and the complementarity of the initial bundle is not fully considered.

Unlike prior bundle generations, we highlight that our Text2Bundle satisfies the following two desirable properties:\emph{Textual-level instruction following}, and \emph{Comprehensive modeling of  personalization and complementarity}. 
As for QBG, it could also be easily compatible and plugged into many search and recommendation scenarios, e.g., as a supplement to the search scenario, or initialization of conversational Bundle recommendation.

\section*{Ethical Considerations}
It is our belief that our proposed novel Query-based Bundle Generation scenario and the Text2Bundle method can mitigate some challenging ethical problems that have been noted in many recommender systems. 
Text2Bundle has the following desirable properties:
\begin{itemize}[leftmargin=*]
\item { }  The ability to give personalized and qualified bundle following user‘s query, speeding up and simplifying the process of information retrieval.
\item { } Users have the opportunity to control their recommendations in a nuanced, or vague, abstract way through language.
\item {} The intention instance generated in the intermediate process provide stronger human interpretability.
\end{itemize}
On the other hand, our proposed system relies on large language models and therefore inherits some well-known problems centered around societal biases, hallucinations, and expensive use of resources. 
We only employ LLM at intention understanding and planning process instead of the whole pipeline, which may alleviate these problems.
Significant further progress needs to be made in areas like debiasing, grounding in factuality and efficient serving before we can safely deploy this type of system in a production setting.


\end{document}